\documentclass[pra,twocolumn,showpacs,floatfix]{revtex4}
\usepackage{graphicx}
\usepackage[usenames]{color} 
\newcommand{\beq}{\begin{equation}}
\newcommand{\eeq}{\end{equation}} 
\newcommand{\beqa}{\begin{eqnarray}}
\newcommand{\eeqa}{\end{eqnarray}} 
\newcommand{\ba}{\begin{array}}
\newcommand{\ea}{\end{array}}

\begin{document}

\title{Localization of a 
Bose-Einstein condensate vortex 
in a bichromatic optical lattice}%and in a random potential
\author{S. K. Adhikari\footnote{adhikari@ift.unesp.br;
URL: www.ift.unesp.br/users/adhikari}}
%\author{L. Salasnich$^{2}$\footnote{luca.salasnich@pd.infn.it; 
%URL: www.padova.infm.it/salasnich}}
\affiliation{Instituto de F\'{\i}sica Te\'orica, UNESP - 
Universidade Estadual Paulista, Barra Funda,  01.140-070 S\~ao Paulo, S\~ao Paulo, Brazil\\
%$^2$CNR-INFM and CNISM, Unit\`a di Padova,
%Dipartimento di Fisica ``Galileo Galilei'', Universit\`a di Padova, Via
%Marzolo 8, 35131 Padova, Italy}
}

\begin{abstract} By numerical simulation of the time-dependent 
Gross-Pitaevskii equation we show that a weakly interacting or 
noninteracting Bose-Einstein condensate (BEC) vortex can be localized 
in a three-dimensional bichromatic quasi-periodic optical-lattice (OL) 
potential generated by the superposition of two standing-wave polarized 
laser beams with incommensurate wavelengths. This is a generalization of the 
localization of a BEC in a one-dimensional bichromatic 
OL 
 as studied in a recent experiment [Roati {\it et al.}, 
Nature {\bf 453}, 895 (2008)]. We demonstrate the stability of the localized state by 
considering its time evolution in the form of a stable breathing 
oscillation in a slightly altered potential for a large period of time. 
{Finally, we  consider the localization of a BEC 
in a random 1D potential in the 
form of several identical repulsive spikes arbitrarily distributed 
in space.
}
 \end{abstract}


\maketitle

\section{Introduction}

The possibility of the 
localization of the electronic wave 
function in a one-dimensional (1D) 
disordered potential as predicted by 
Anderson in his pioneering work  \cite{anderson}
has drawn the attraction of researchers in different areas. 
Different forms of localization have been observed experimentally in 
diverse contexts, such as in   electromagnetic 
waves \cite{light,micro}, in sound waves \cite{sound}, 
and also more recently 
in quantum matter waves \cite{billy,roati,chabe,edwards}. 
Using a cigar-shaped noninteracting 
Bose-Einstein condensate (BEC) \cite{book1,book2}
of $^{87}$Rb atoms, 
Billy {\it et  al.} \cite{billy} demonstrated its  exponential localization  
when released into a 1D waveguide in the 
presence of a controlled disorder created by a laser speckle. 
In another experiment, 
using a  cigar-shaped   noninteracting
BEC of $^{39}$K atoms, 
Roati {\it et  al.} \cite{roati} demonstrated its  localization 
in a 1D bichromatic optical-lattice (OL) 
potential created by the superposition of 
two standing-wave polarized laser beams with different
wavelengths. 
The noninteracting BEC of 
$^{39}$K atoms was created \cite{roati} by tuning the inter-atomic 
scattering length to zero near a Feshbach resonance \cite{fesh}. 

The localization of a BEC in a 1D bichromatic quasi-periodic OL potential and related 
topics have been the subject matter of several theoretical 
\cite{das,dnlse,harper,boers,aubry,thouless,random,optical,modugno,modugno2,roux,roscilde,paul,adhikari} 
 and experimental \cite{billy,roati,chabe,edwards}
studies. After the pioneering experiments 
\cite{roati,billy} on the localization of a 1D cigar-shaped BEC, a 
natural extension of this phenomenon would be to achieve localization in 
higher dimensions, e.g., in two (2D) and three (3D) dimensions.  We 
address this important issue in the present investigation. Using the 
complete numerical solution of the Gross-Pitaevskii (GP) equation 
\cite{GP}, here we study the localization of a (nonrotating) 
disk-shaped  BEC in 2D and also a BEC in 3D, with a small 
nonlinearity, in the presence of bichromatic OL potentials along 
orthogonal directions. (For zero nonlinearity the problem of a  stationary 
state of a $d$-dimensional $-$ $d=2,3$ $-$ nonrotating BEC trivially 
decouples into $d$ 1D problems and for a large repulsive nonlinearity 
the localization is destroyed \cite{adhikari,modugno2}.) In 3D 
another nontrivial phenomenon can happen, i.e., the generation of a stable  
vortex state of unit quantum.  
Here we demonstrate the localization of a stable nontrivial 
vortex state in a 3D BEC under the action of  bichromatic OL 
potential along the axial direction and harmonic potentials along 
radial directions.  We 
find, as in 1D \cite{adhikari,modugno2}, a repulsive nonlinearity has a 
strong effect on localization and a not-too-large nonlinearity destroys 
localization in all cases. We exhibit results for zero and small 
nonlinearities. (Effects of a weak nonlinearity in Anderson 
localization have been shown experimentally in light waves in photonic 
crystals \cite{roati,lahini}.)

{ In the presence of strong disorder, the localized state 
could be quite similar to a localized state of Gaussian shape in an 
infinite potential. However, the more interesting case of localization 
is in the presence of a weak disorder when the system is localized due 
to the quasi-periodic nature of the potential \cite{billy,roati} and not 
due to the strength of the lattice. When this happens the localized 
state acquires an exponential tail. The present localization with 
an exponential tail
in a 
quasi-periodic OL potential with a deterministic weak 
disorder is a special case of
Anderson localization in a fully disordered potential and is well described 
in the Aubry-Andr\'e model \cite{aubry}.  }

{{In the pioneering study, Anderson considered localization of 
electron(s) in a 1D random potential generated by impurity or disorder 
arbitrarily distributed on a lattice. This is of interest in the study 
of a BEC and we consider its localization in a disorder potential in the 
form of identical narrow repulsive spikes (simulating delta functions) 
distributed randomly in space. (This is different from a speckle 
potential where the spikes also have different strength and width.) We 
find that as small as four such spikes can localize a noninteracting 
 BEC in 1D. 
This is not just of academic interest as such potential can be 
created in 1D and possibly in 2D for a BEC in an atom chip 
\cite{chip,chip2}.  A single repulsive spike separating the BEC into two 
parts is now routinely created \cite{chip2}.}}

There have been theoretical studies on different aspects of Anderson 
localization which are worth mentioning \cite{adhikari}. 
Wobst {\it et 
al.} \cite{wobst} considered Anderson localization in higher dimensions. 
%The relation of Anderson localization with Heisenberg uncertainly 
%principle has also been addressed \cite{wobst2}.  
Sanchez-Palencia {\it 
et al.} and Cl\'ement {\it et al.} considered Anderson localization in a 
random potential \cite{random,stoca,optical}. Damski {\it et al.} and Schulte {\it et 
al.} considered Anderson localization in disordered OL potential 
\cite{optical}. There have been studies of Anderson localization with 
other types of disorder \cite{other}. Effect of interaction on Anderson 
localization was also studied \cite{interaction,dnlse}. Anderson 
localization in BEC under the action of a disordered potential in 2D and 
3D has also been investigated \cite{2D3D}.

%The spatial ordering of a bichromatic OL potential is intermediate 
%between deterministic and random disorders \cite{harper,aubry,thouless}.  
%(Anderson localization can also be achieved in the case of random 
%disorder \cite{stoca}.) In particular, the 1D discrete Aubry-Andre model 
%of quasi-periodic confinement \cite{aubry,thouless,das} displays a 
%transition from extended to localized states which resembles the 
%Anderson localization in random disorder \cite{random,optical}. 

In Sec. \ref{II} we present a brief account of the nonlinear 2D and 3D 
time-dependent GP equations used in our study and of the variational 
solution of the same under appropriate conditions. The generalization of 
the equation to study vortex states in 3D is also presented.  In Sec. 
\ref{III} we present numerical results of localization employing
 time propagation using the semi-implicit Crank-Nicolson 
algorithm. The wave function of the localized states has a central Gaussian 
(variational) form with a long exponential tail. 
 First we consider in Sec. \ref{IIIA}
the 
localization of a 2D disk-shaped BEC.
 We also consider localization of a 3D BEC and a 3D 
vortex BEC in  bichromatic OL potential(s) in Sec. \ref{IIIB}. 
{{In Sec. \ref{IIIC} we consider 
the localization of a noninteracting BEC 
 for a random potential comprised of arbitrarily 
distributed narrow spikes. }} 
 In Sec. \ref{V} we present a 
brief discussion and concluding remarks.

\section{Theoretical formulation of localization}

\label{II}

The quasi-periodic bichromatic OL potentials generated by two 
standing-wave polarized laser beams of incommensurate  wavelengths
in the $x$ direction 
have the following generic forms \cite{roati}:
\begin{eqnarray}
\label{pot1}
U(  x)=\sum_{i=1}^2 s_iE_{i}\cos^2(k_i  x),\\
U(  x)=\sum_{i=1}^2 s_iE_{i}\sin^2(k_i  x),
\label{pot2}
\end{eqnarray}
where $s_i, i=1,2,$ are the amplitudes  of the OL potentials in units of 
respective recoil energies $E_i=2\pi^2 \hbar^2/(m  \lambda_i^2)$, and 
$k_i=2\pi/ \lambda_i$, $i=1,2$ are the respective wave numbers and
$\lambda_i$ are the wave lengths,
$\hbar(\equiv h/2\pi) $ 
is the reduced Planck constant, and $m$ the mass of an atom. 

If we have  a single periodic potential of forms (\ref{pot1}) and (\ref{pot2}) 
with $s_2=0,$  the  solution of the 
Schr\"odinger equation cannot be localized. One can have localization 
if a second periodic component with a different 
frequency is introduced 
in Eqs. (\ref{pot1}) and  (\ref{pot2}). 
These  localized states  are not
 the gap solitons, which are localized states  
in the solution of a nonlinear Schr\"odinger equation with a repulsive 
nonlinearity 
appearing 
in the band-gap of the spectrum of the linear Schr\"odinger 
equation \cite{gs}. 
%These gap solitons with finite spatial extension 
%are {excited  states} of the system { without a linear 
%counterpart}, whereas the present localization is achieved with the linear 
%Schr\"odinger equation.  

%Experimentally, the atom-atom interaction in a
%BEC can be made  zero by varying an external magnetic field near a Feshbach 
%resonance. This gives the 
%unique opportunity to study the localization in a  BEC composed of ideal
%bosonic  atoms without interaction.    
The BEC in 3D is described by the GP equation 
\begin{eqnarray}\label{gp}
i\hbar\frac{\partial \phi({\bf r},\tau)}{\partial \tau}
=\left[ -\frac{\hbar^2\nabla^2}{2m}+V({\bf r})+g|\phi({\bf r},\tau)|^2
\right] \phi({\bf r},\tau),
\end{eqnarray}
where $g=4\pi\hbar^2aN/m, \int |\phi({\bf r},\tau)|^2 d{\bf r}=1$, 
$\tau$ the time, 
$N$ the number of atoms, $V({\bf r})$ is the  
trap, and $a$ is the atomic scattering length. 
With three quasi-periodic, bichromatic OL potential in $x,y$, and $z$ 
directions, after canceling the factor $\hbar^2/m$ from both sides of Eq. 
(\ref{gp}),
the GP equation in 
explicit notation becomes 
\begin{eqnarray}\label{gp3d}
i\frac{\partial \phi(x,y,z,t)}{\partial t}
&=&\biggr[ -\frac{1}{2}\left(\partial_x^2+ \partial_y^2+\partial_z^2  
\right)+V(x)+V(y) \nonumber \\ &+&V(z)+ g|\phi(x,y,z,t)|^2
\biggr] \phi(x,y,z,t),
\end{eqnarray}
where $g=4\pi a N$, $\partial_x$'s denote space derivatives,  and 
time $t\equiv \tau \hbar/m $ is now expressed in units of $m/\hbar$.
Note that Eq. (\ref{gp3d}) is not expressed in dimensionless units.
The variables $x,y,z,\lambda_i$ are in actual units of length ($L$), 
$|\phi|^2$ is in units 
of $L^{-3}$ with normalization 
 $\int_{-\infty}^{\infty} \int_{-\infty}^{\infty} \int 
_{-\infty}^{\infty}|\phi(x,y,z,t)|^2 dx dy dz =1$. 
In Eq. (\ref{gp3d}) the scaled potentials $V(x)\equiv U(x)m/\hbar^2$ 
are now 
defined by one of the two following expressions:
\begin{eqnarray}
V(x)&=&\sum_{i=1}^2\frac{2\pi^2s_i}{\lambda_i^2}\cos^2\left(\frac{2\pi x}{\lambda_i}\right),
\label{p1}
\\
\label{p2} 
V(x)&=&\sum_{i=1}^2\frac{2\pi^2s_i}{\lambda_i^2}\sin^2\left(\frac{2\pi x}{\lambda_i}\right).
\end{eqnarray}
In case of a 2D and 3D BEC, instead of having the same potential, e.g., 
 (\ref{p1}) or (\ref{p2}) in different directions, one can choose different 
potentials along different directions.

For axially-symmetric traps, the GP equation can be easily generalized 
to include a vortex state, as shown in Ref. \cite{ds,book2}, in 
axially-symmetric coordinates ${\bf r}\equiv (\rho,z),$ where $\rho$ is 
the radial coordinate and $z$ the axial coordinate. To obtain a singly 
quantized vortex state of angular momentum $\hbar$ around $z$ axis, 
one has to explicitly introduce a phase (equal to the azimuthal angle) in 
the wave function. (Vortex states of higher angular momentum are unstable and 
decays into multiple states of angular momentum $\hbar$.) This procedure 
introduces a centrifugal term in the GP equation representing a vortex state
\cite{ds,book2}.
Thus we can 
study a localized quantized BEC vortex of unit angular momentum in the 
axially-symmetric potential, where the bichromatic OL potential $V(z)$ 
is placed along the axial $z$ direction and an harmonic trap $\rho^2/2$ 
along the transverse radial $\rho$ direction.  The modified GP equation 
for such a vortex is given by \cite{ds} \begin{eqnarray}\label{gp3dv} 
&&i\frac{\partial \phi(\rho,z,t)}{\partial t} =\biggr[ 
-\frac{1}{2}\frac{\partial^2}{\partial \rho^2}-\frac{1}{2\rho} 
\frac{\partial}{\partial \rho}-\frac{1}{2}\frac{\partial^2}{\partial 
z^2} +\frac{1}{2\rho^2} \nonumber \\ &&+\frac{1}{2} 
\rho^2+V(z)+g|\phi(\rho,z,t)|^2\biggr] \phi(\rho,z,t), \end{eqnarray}
 where we have    
explicitly included the  angular momentum dependent 
centrifugal term ${1}/({2\rho^2})$ \cite{ajp}. 
Because of the centrifugal term, the 
density of the vortex  should be zero along the $z$ axis.

If there is a strong harmonic trap in the axial $z$ direction, one can derive a
reduced quasi-2D GP equation for a disk-shaped BEC, 
which can be written in dimensionless harmonic oscillator units as \cite{CPC}
\begin{eqnarray}\label{gp2d}
i\frac{\partial \phi(x,y,t)}{\partial t}
&=&\biggr[ -\frac{1}{2}\left(\partial_x^2+ \partial_y^2 
\right)+V(x)+V(y) \nonumber \\ &+& g|\phi(x,y,t)|^2
\biggr] \phi(x,y,t),
\end{eqnarray}
 where $g=2\sqrt{2\pi} a N$ and $V(x)$ is given by Eqs. (\ref{p1}) or 
(\ref{p2}). In Eq. (\ref{gp2d}) the normalization is $\int_{-\infty}^
{\infty} \int _{-\infty}^{\infty}|\phi(x,y,t)|^2 dx dy =1$.
In Eq. (\ref{gp2d}) lengths are expressed in units of $l_z\equiv
\sqrt{\hbar/m\omega_z}$ with $\omega_z$ the frequency in $z$ 
direction, 
$\phi(x,y,t)$ in units of $l_z^{-2} $ 
and time in units of $\omega_z^{-1}$.

Finally, for the sake of completeness we note that if there is a strong 
harmonic trap in transverse $y$ and $z$ directions, one can derive a 
reduced quasi-1D GP equation for a cigar-shaped BEC, which 
can be written in dimensionless harmonic oscillator units as \cite{CPC} 
\begin{eqnarray}\label{gp1d} i\frac{\partial \phi(x,t)}{\partial t} 
=\biggr[ -\frac{1}{2}\partial_x^2 +V(x) + g|\phi(x,t)|^2 \biggr] 
\phi(x,t), \end{eqnarray} where $g=2a N$ and $V(x)$ is given by Eqs. 
(\ref{p1}) or (\ref{p2}). Now the normalization is $ 
\int_{-\infty}^{\infty}|\phi(x,t)|^2 dx =1$.  Equation (\ref{gp1d}) has 
been used \cite{modugno,adhikari}
for the study of localization in 1D in bichromatic 
OL potential. In Eq. (\ref{gp1d}) lengths are expressed in units of $l\equiv
\sqrt{\hbar/m\omega_\perp}$ with $\omega_\perp$ the frequency in transverse 
$y$ and $z$ 
directions, 
$\phi(x,t)$ in units of $l^{-1} $ 
and time in units of $\omega_\perp^{-1}$. 

%In Ref. \cite{adhikari} a complete numerical solution of 
%(\ref{gp1d}) was used in the study, whereas in Ref. \cite{modugno}
%an approximate solution of Eq. (\ref{gp1d}) was performed using  
%a model discrete nonlinear Schr\"odinger equation \cite{dnlse,modugno}.

Although we use potentials (\ref{p1}) and (\ref{p2}) in our study, there is 
some difference between these two potentials.  
Potential (\ref{p2}) generates a different type of
localized states
compared to potential  (\ref{p1}). Potential  (\ref{p2}) has a local
minimum at the center, consequently  stationary solutions
with this potential have a maximum there. However, potential
(\ref{p1}) has a local maximum at the center  corresponding to a minimum of
the stationary solution.

Usually the stationary localized  states formed with quasi-periodic OL
potentials (\ref{p1}) and (\ref{p2}) occupy many sites of the
quasi-periodic OL potential and have many local maxima and minima. 
For
certain values of the parameters, potential (\ref{p2}) leads to
localized states confined practically to the central cell  of the
quasi-periodic OL potential. When this happens, a variational
approximation
with Gaussian ansatz leads to a reasonable prediction for the localized 
state in the central region. (However, it has an exponential tail at large 
distances.)

To derive a simple  variational solution of linear 
Eqs. (\ref{gp3d}), (\ref{gp2d}), 
and (\ref{gp1d}) in 3D, 2D, and 1D,
respectively, in a unified fashion,  
we adopt the convenient notation 
${\bf r}\equiv (x,y,z)$ in 3D, $\equiv (x,y)$ in 2D, and 
$\equiv (x)$ in 1D.
The stationary form of the linear 
Schr\"odinger equations  (\ref{gp3d}), (\ref{gp2d}),
and (\ref{gp1d})
(with
$i\partial /\partial t$ replaced by a chemical potential 
$\mu$)
with
potential (\ref{p2}) can
be derived from the
following Lagrangian
\begin{eqnarray}\label{lag}
L&=&\int_{-\infty}^\infty \biggr[ \mu |\phi({\bf r})|^2
-\frac{1}{2}|\nabla \phi({\bf r})|^2
\nonumber \\
&-&  V({\bf r})|\phi({\bf r})|^2-\frac{g}{2}
|\phi({\bf r})|^4
\biggr]
d{\bf r} -\mu,
\end{eqnarray}
by demanding $\delta L/\delta \phi = \delta L/\delta \mu=0$. 
To apply the variational
approximation we use the
Gaussian ansatz \cite{PG}
\begin{equation}\label{ans}
\phi({\bf r})=\left(\frac{\pi^{-1/4}}{\sqrt w}\right)^d
{\sqrt {\cal N}}\exp\left(- \frac{\sum_{j=1}^d x_j^2}{2w^2}
\right),
\end{equation}
where $d$ is the dimension of space, $x_1\equiv x, x_2\equiv y,
x_3\equiv z,$ 
and the  
variational parameters are the norm $\cal N$, width $w$, and 
$\mu$.
This ansatz implies that the center of the stationary state  is placed
at the local
minimum at $x_i=0, i=1,2,3$ 
of the quasi-periodic OL potential. The
substitution
of ansatz (\ref{ans}) in Lagrangian (\ref{lag}) leads to
\begin{eqnarray}\label{lag2}
L&=&\mu({\cal N}-1)-\frac{d{\cal N}}{4w^2}
+d{\cal N}\sum_{i=1}^2\frac{{A_i 
 }}{2}[
\exp(-\alpha_i^2
w^2)-1]\nonumber
 \\&-&\frac{g{\cal N}^2}{2} 
\left( \frac{1}{\sqrt{2\pi}w}  \right)^d,
\end{eqnarray}
where $A_i=2\pi^2s_i/\lambda_i^2, \alpha_i=2\pi/\lambda_i.$
The first variational equation from Eq. (\ref{lag2}),
$\partial L/\partial \mu=0,$  yields ${\cal N}=1$, which will be used
in other variational equations. The second variational equation
$\partial L/\partial w=0,$  yields
\begin{equation}\label{cf}
1=\sum_{i=1}^2 2\alpha_i^2 A_i w^4 \exp(-\alpha_i ^2 w^2)
-\frac{g}{(2\pi)^{d/2}w^{d-2}}
,
\end{equation}
  and determines the width $w$.
The last variational equation $\partial L/\partial {\cal N}=0,$ yields
\begin{equation}\label{energy}
\mu = \frac{d}{4w^2}-\sum_{i=1}^2\frac{{dA_i}}{2}[\exp(-\alpha_i^2 w^2)-1]
+g\left( \frac{1}{\sqrt{2\pi}w}   \right)^d
,
\end{equation}
 which determines the chemical potential. 

{ In addition to considering a disorder potential in the form of a 
bichromatic lattice, we also consider the following disorder potential 
in the form of randomly distributed $S$ repulsive identical Gaussian 
spikes in 1D along the $x$ axis \begin{eqnarray}\label{randpot} V(x) = 
\sum_{i=1}^S B_i \exp[-c_i (x-\beta_{i})^2], \end{eqnarray} where $B_i$ is the 
amplitude of the Gaussian spike, $c_i$ is its width, and $\beta_{i}$ is its 
random position. If the spikes are placed at constant periodic spacing, 
no localization can be obtained. There will be localization if the 
positions $\beta_{i}$ are random. In 2D and 3D the appropriate potentials 
are $V(x)+V(y)$ and $V(x)+V(y)+V(z).$ We study the localization of a BEC 
in random potential (\ref{randpot}).

For an analytical understanding of the problem, next we present a 
variational analysis. As the potential (\ref{randpot}) is Gaussian
a variational analysis based on the Gausssian ansatz is fully integrable. 
Only the potential term in the Lagrangian gets modified 
and in this case the third term in the Lagrangian of Eq. (\ref{lag2})
becomes
\begin{equation}
-d{\cal N}\sum_{i=1}^SB_i\frac{\exp\left[ -c_i \beta_{i}^2/\gamma_i 
\right]}{\sqrt{\gamma_i}},
\end{equation}
and Eq. (\ref{cf}) for width gets modified into
\begin{equation}\label{cf2}
1=\sum_{i=1}^S\frac{2c_iB_iw^4[2c_i\beta_i^2-\gamma_i]}
{\exp\left[ c_i \beta_{i}^2/\gamma_i
\right]\sqrt{\gamma_i}}-\frac{g}{(2\pi)^{d/2}w^{d-2}},
\end{equation}
where $\gamma_i=1+c_iw^2$. The expression for chemical potential becomes
\begin{equation}\label{energy2}
\mu = \frac{d}{4w^2}+\sum_{i=1}^S\frac{dB_i\exp(-c_i\beta_i^2/\gamma_i)}{\sqrt{\gamma_i}}
+g\left( \frac{1}{\sqrt{2\pi}w}   \right)^d
.
\end{equation}
}

%In all cases the localized states for $g=0$ can have a quasi-Gaussian 
%form and these states can be easily represented by the variational analysis. 
%The states for $g>0$ develop ondulating tails, which cannot be analysed 
%by the Gaussian variational ansatz. Hence we apply the variational 
%analysis to only the localized states with $g=0$ with a quasi-Gaussian form. }

\section{Numerical Results}

\label{III}

We performed numerical simulation employing imaginary- and real-time 
propagations with Crank-Nicolson discretization scheme \cite{bo,CPC} 
using adequately small space and time steps necessary for obtaining 
converged solutions. In practice we used space and time steps smaller 
than 0.025 and 0.0005, respectively, and sometimes as small as 0.0025 
and 0.00002, respectively.  We use the FORTRAN programs provided in Ref. 
\cite{CPC} for our purpose. 
%Imaginary-time propagation routine can 
%determine the strongly localized state confined to a small number of OL 
%sites (preferably a single site) in an efficient fashion. 
For 
checking the consistency of our calculation we compared our real-time 
results with imaginary-time results and we verified that the two sets of 
results were in agreement with each other. Because of the oscillating 
nature of the bichromatic OL potential, great care was needed to obtain 
a { localized state} precisely. The accuracy of the numerical simulation 
was tested by varying the space and time steps as well as the total 
number of space and time steps. Although we used the time-dependent GP 
equation for the study of localization, all results reported in this 
paper, except those in Fig. \ref{fig4}, 
are stationary results independent of time.
In  Fig. \ref{fig4} we study time-dependent stability dynamics of the 
localized states obtained with real-time propagation.

\begin{figure}%[!ht]
\begin{center}
\includegraphics[width=\linewidth]{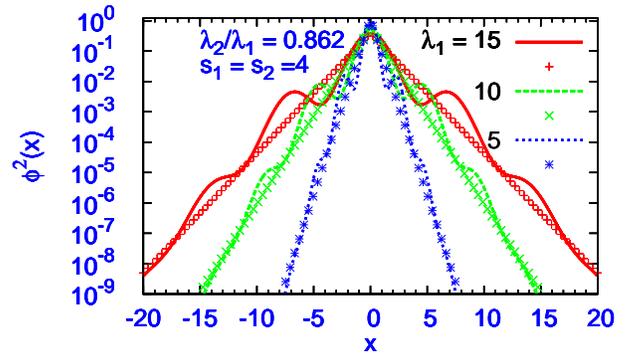}
\end{center}

\caption{(Color online) Numerical density $\phi^2(x)$ (line) and 
its exponential fit (symbol)  vs. $x$ for $\lambda_1=15,10$ and 5.  
}  
\label{fig1}
\end{figure}

\begin{figure}%[!ht]
\begin{center}
\includegraphics[width=\linewidth]{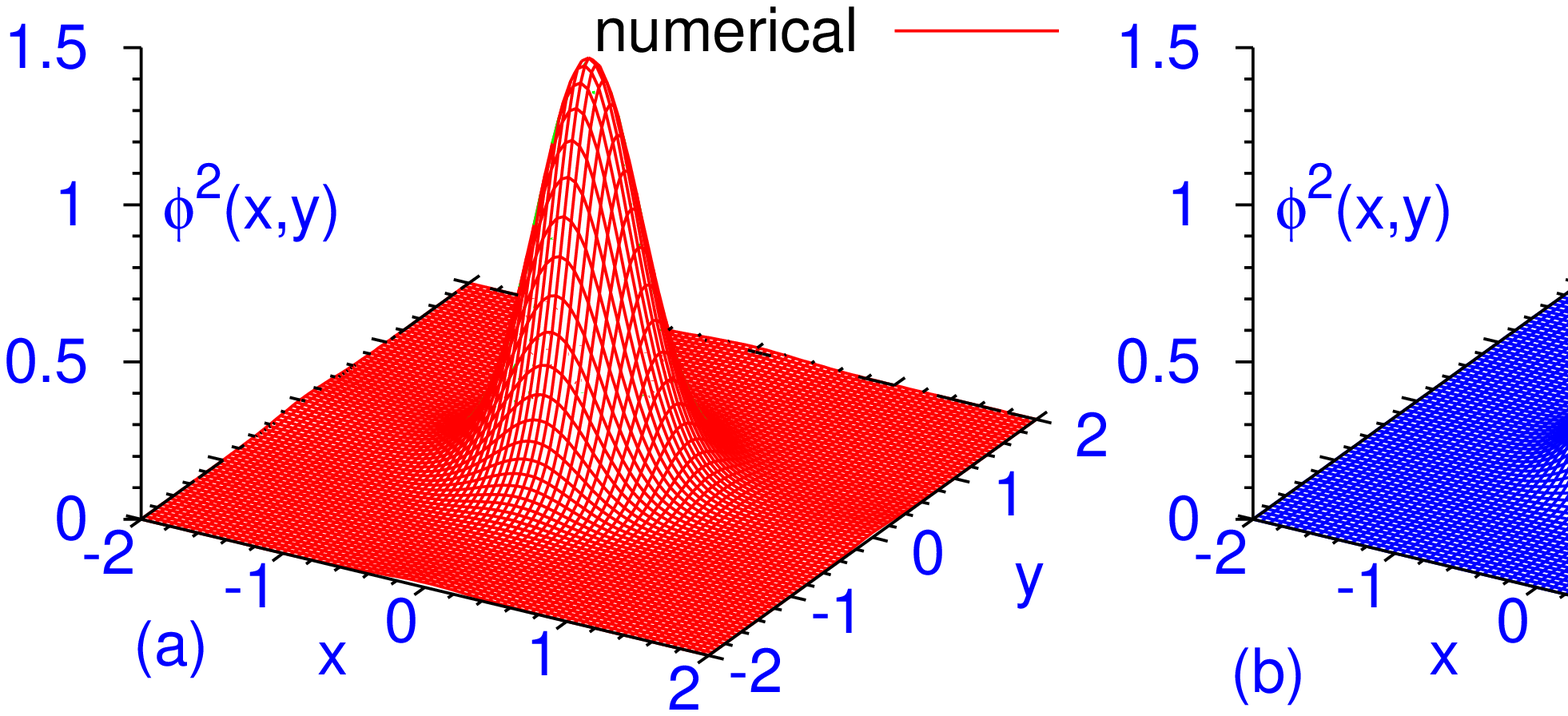}
\includegraphics[width=.49\linewidth]{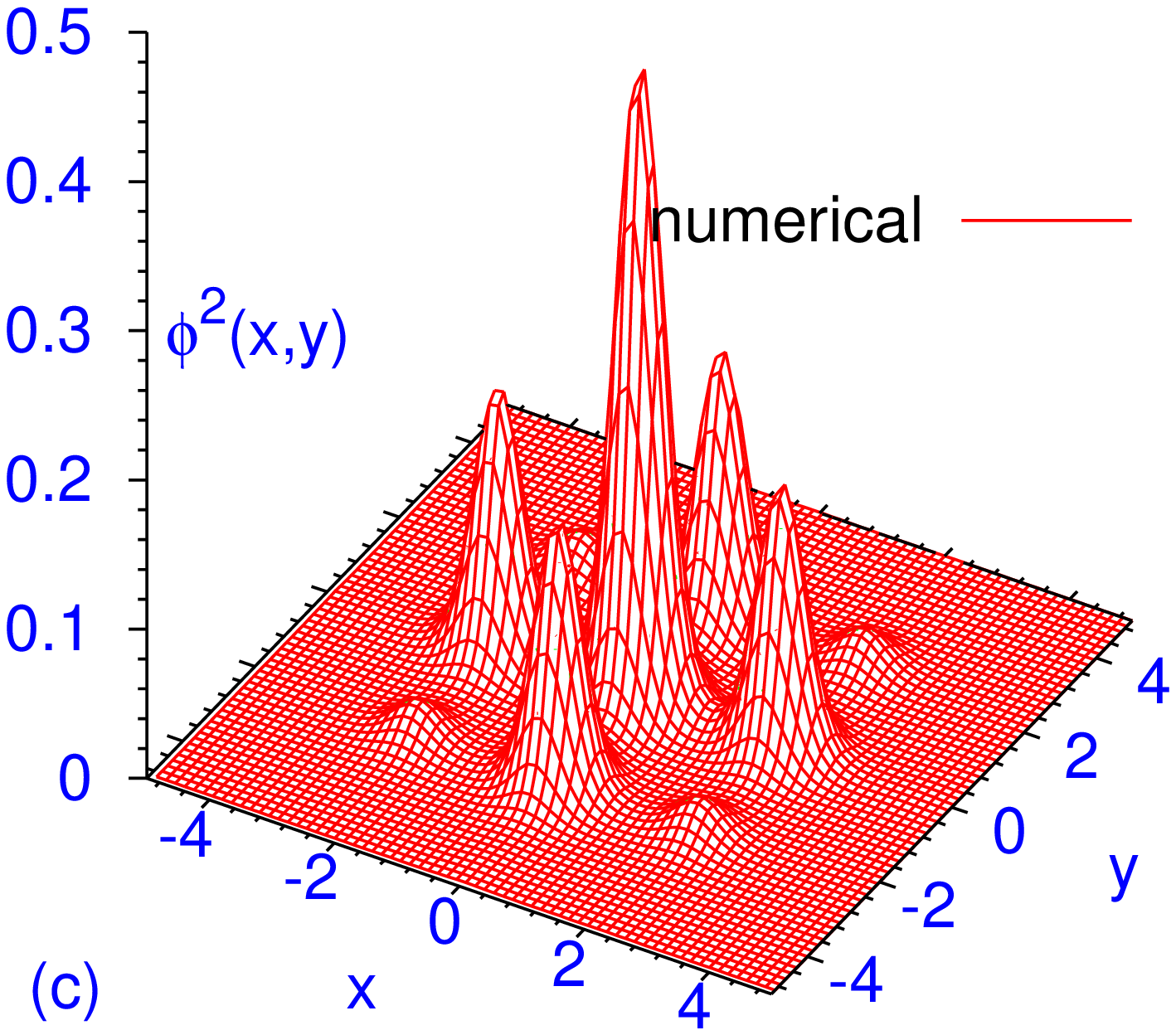}
\includegraphics[width=.49\linewidth]{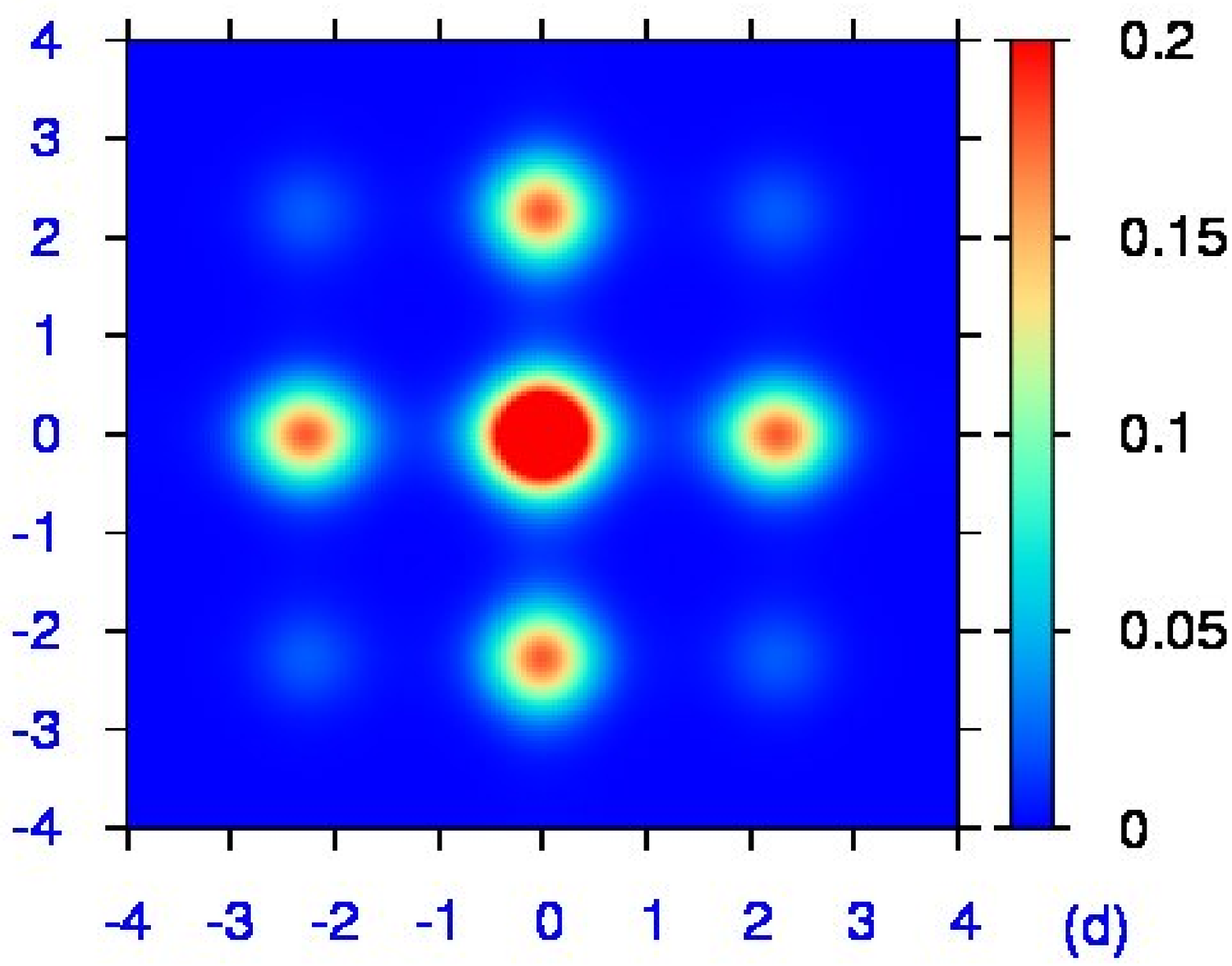}
\includegraphics[width=.49\linewidth]{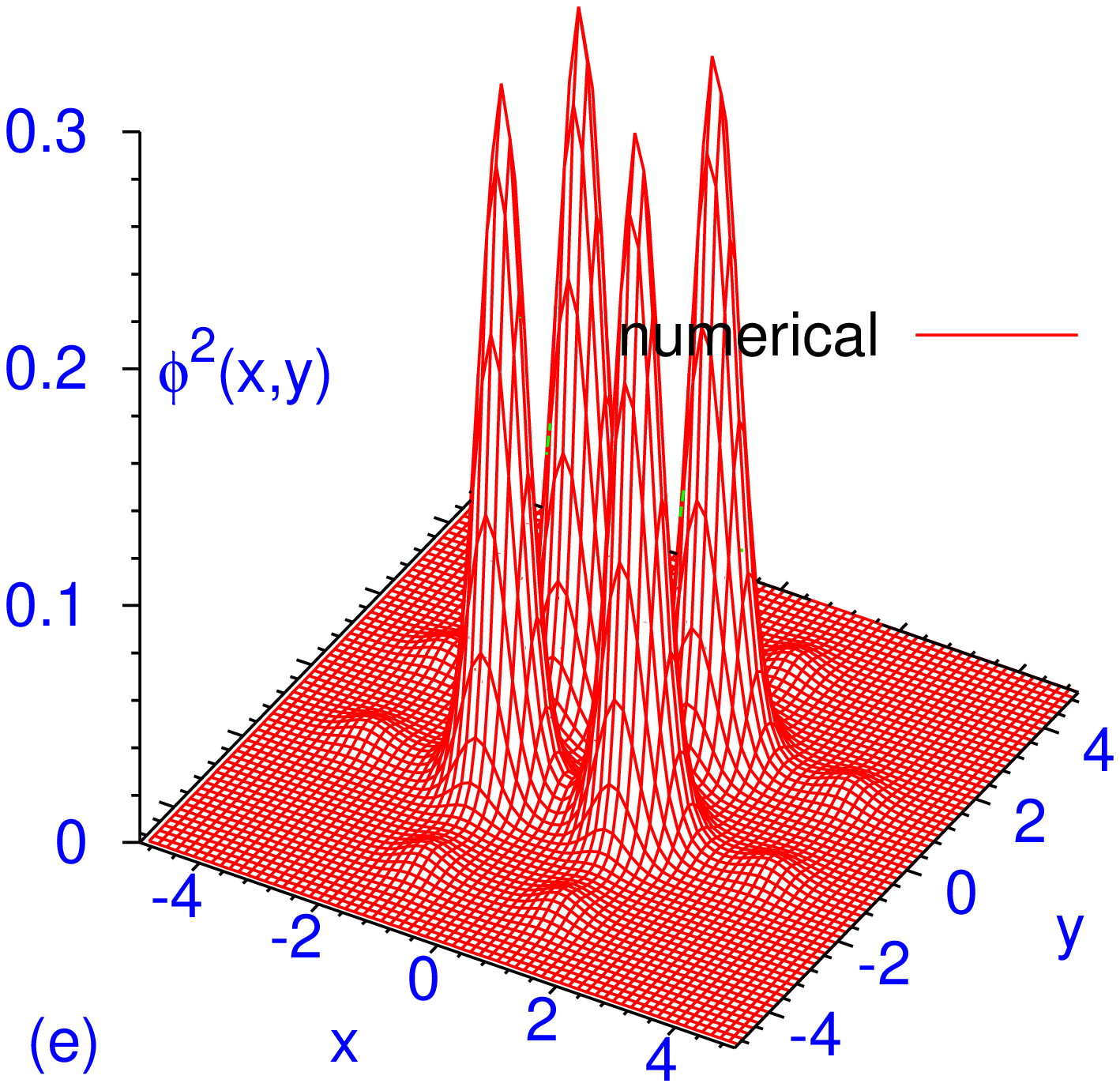}
\includegraphics[width=.49\linewidth]{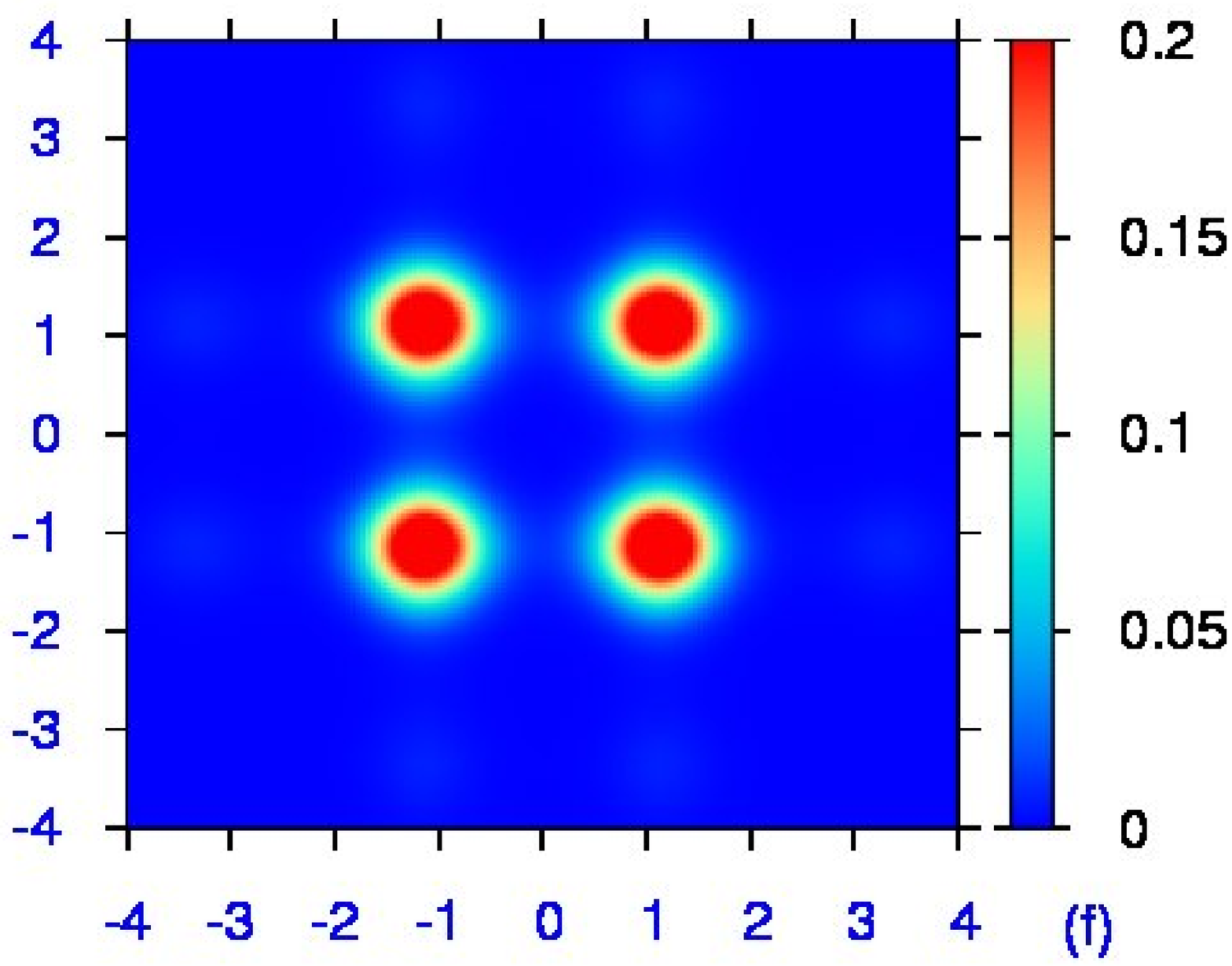}
\end{center}

\caption{(Color online) (a) Numerical and (b) variational   density  
$\phi^2(x,y)$ vs.
$x$ and $y$  from a solution of Eqs. (\ref{gp2d}) and  (\ref{p2})
for a disk-shaped BEC for  $g=0$. 
(c) Numerical density
$\phi^2(x,y)$ vs.
$x$ and $y$  and (d) its contour plot 
from a solution of Eqs. (\ref{gp2d}) and  (\ref{p2})
for a disk-shaped BEC for  $g=2$.
(e) Numerical density
$\phi^2(x,y)$ vs.
$x$ and $y$  and (f) its contour plot 
from a solution of Eqs. (\ref{gp2d}) and  (\ref{p1})
for a disk-shaped BEC for  $g=2$.
Quantities   $\phi^2(x,y)$ and $x,y$ are all   in dimensionless 
harmonic oscillator  units.}
\label{fig2}
\end{figure}

\subsection{2D Bichromatic Optical Lattice }

\label{IIIA}

To study the localization of a BEC in 2D and 3D with potentials 
(\ref{p1}) and (\ref{p2}), we set the ratio $\lambda_2/\lambda _1 =0.86$ 
(roughly the same ratio $\lambda_2/\lambda_1$ as in the experiment of 
Roati et al. \cite{roati}).
First we consider the solution of  Eq. (\ref{gp2d}) 
 for a disk-shaped BEC. 
To understand the nature of these localized states, 
we 
consider the localized states with larger values of $\lambda_1$.  Such 
states with a {\it large} $s_2/s_1 (=1)$ occupy a small number of OL sites 
and hence their numerical
simulation can be performed relatively easily. All results reported in 
Secs. \ref{IIIA} and \ref{IIIB} are obtained with the following parameters
in the bichromatic OL potentials (\ref{p1}) and (\ref{p2}): 
$\lambda_1=5,$ $\lambda_2/\lambda_1=0.86,$ and $ s_1=s_2=4$.
{
It now remains to be seen if with this set of parameters the localized state 
is in the limit of weak disorder with an exponential tail. In 1D this would 
mean $\phi(x) \sim \exp(-x/L_{\mathrm{loc}})$, where $L_{\mathrm{loc}}$ is the 
localization length. In Fig. \ref{fig1} we plotted the numerical probability 
$\phi^2(x)$ vs. $x$ on a log scale, together with a fitting exponential 
function with $L_{\mathrm{loc}}=0.75, 1.5 $ and 2.2, respectively,  
for $\lambda_1=5,10$ and 15. These localization lengths  are large
compared to the 
root mean square (rms) size of the localized states, which 
are 0.53, 1.05, and 1.55 for $\lambda_1=5,10$ and 15.  
%From this  the exponential fit with large localization length is obvious.   
The wave functions shown in Fig. \ref{fig1} are identical to those shown 
in Fig. 1 (a) of Ref. \cite{adhikari} where the central part of these wave 
functions are well fitted to  Gaussian variational approximations. 
}

\begin{figure}%[!ht]
\begin{center}
\includegraphics[width=.49\linewidth]{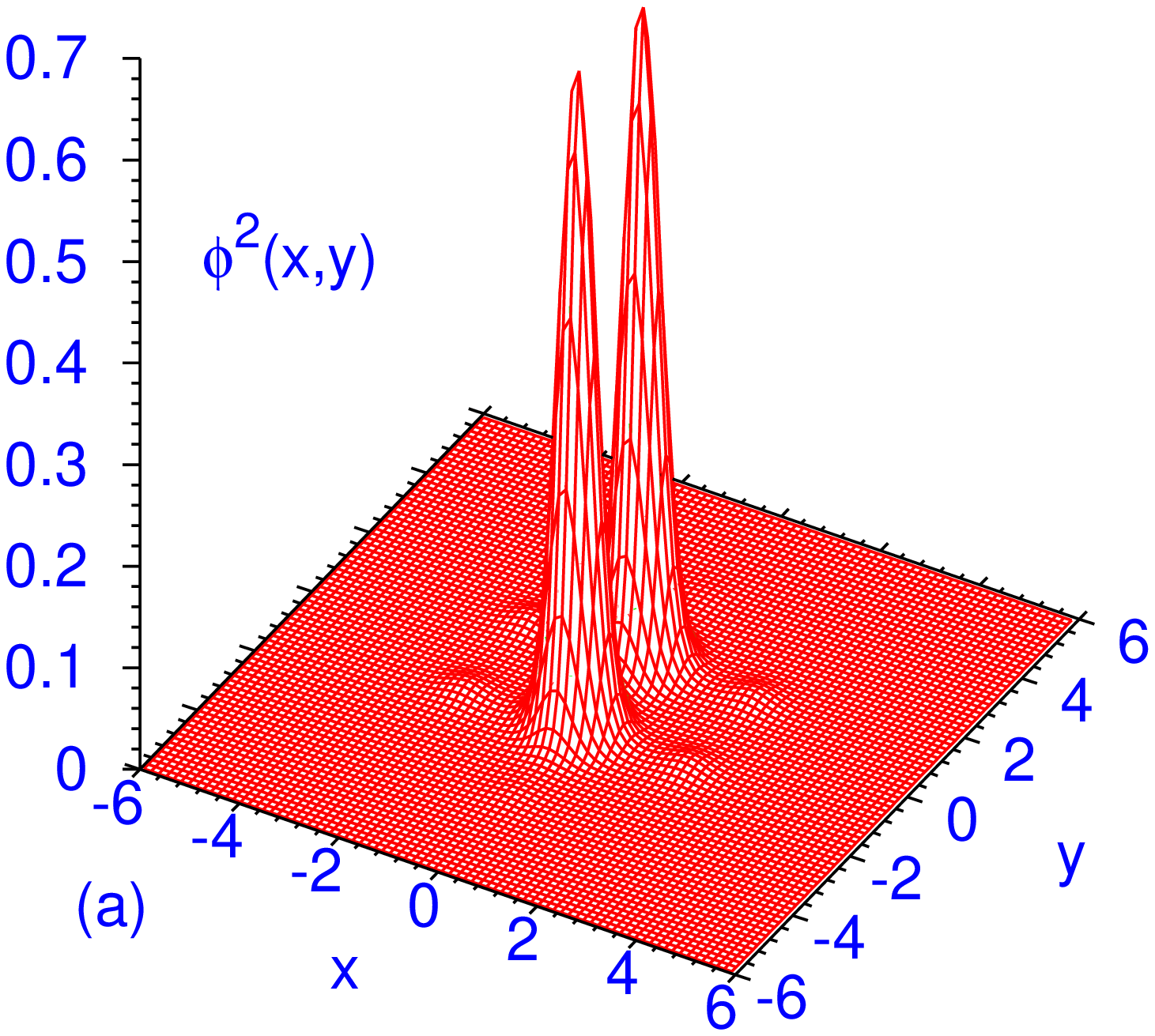}
\includegraphics[width=.49\linewidth]{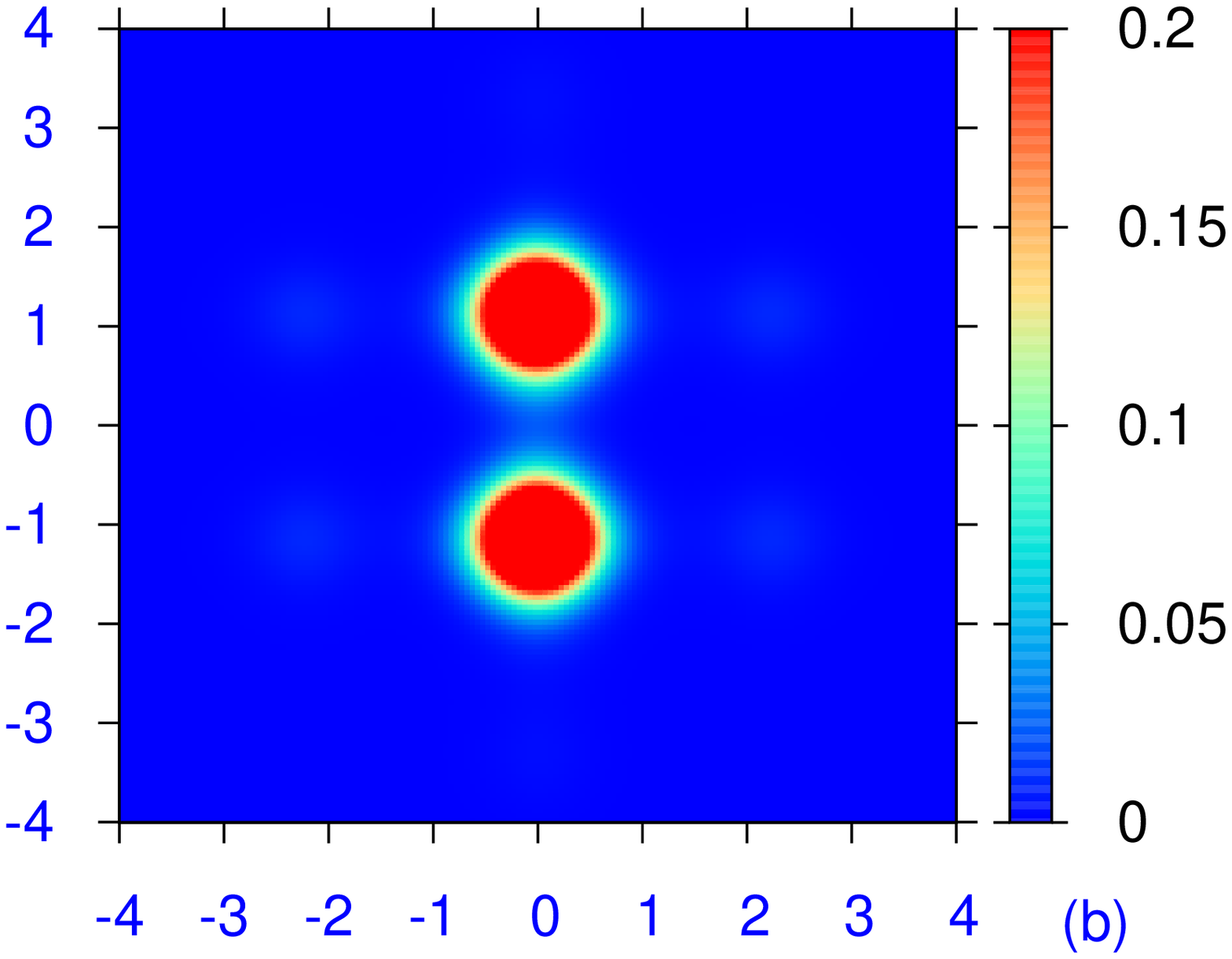}
\includegraphics[width=.49\linewidth]{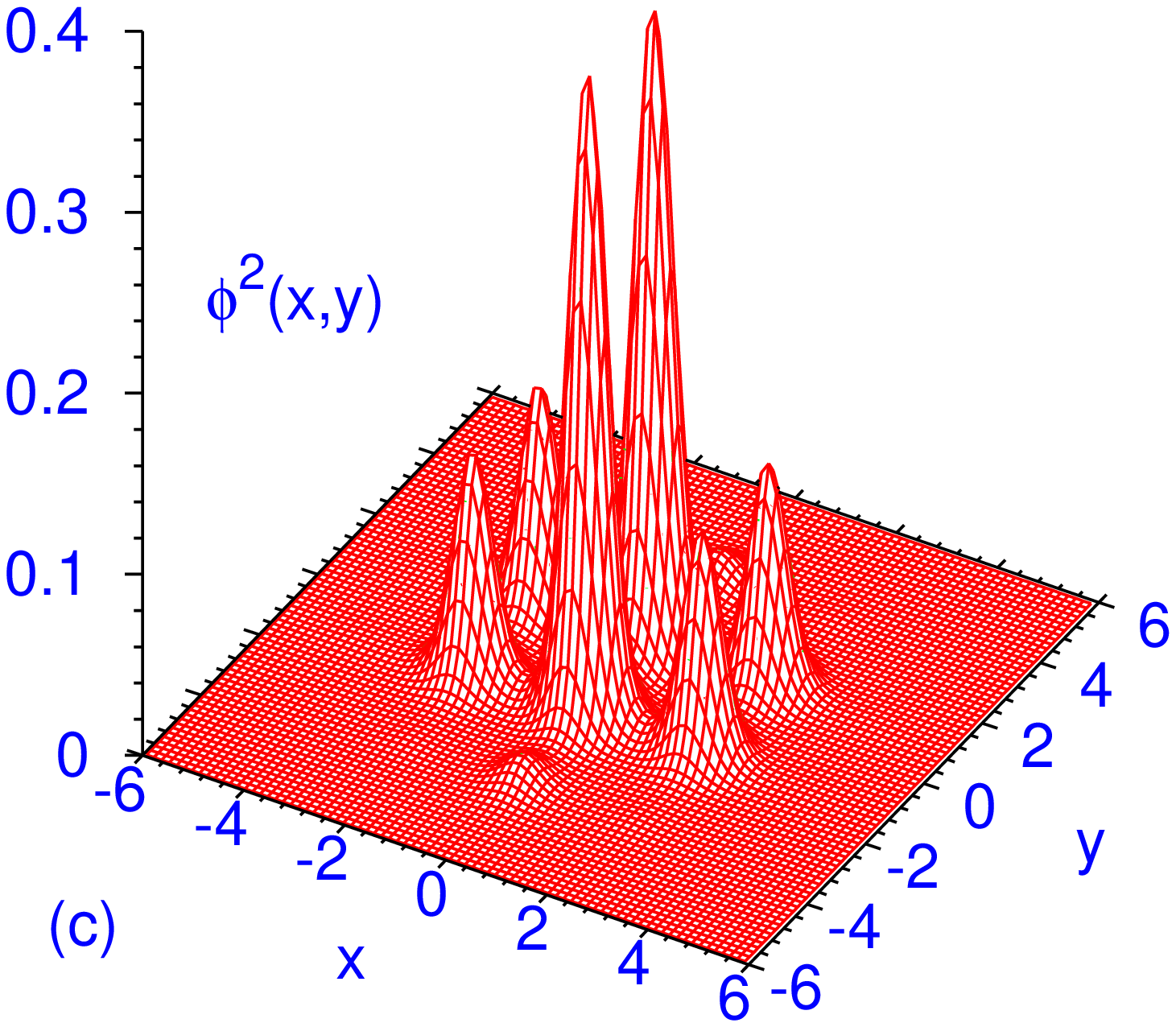}
\includegraphics[width=.49\linewidth]{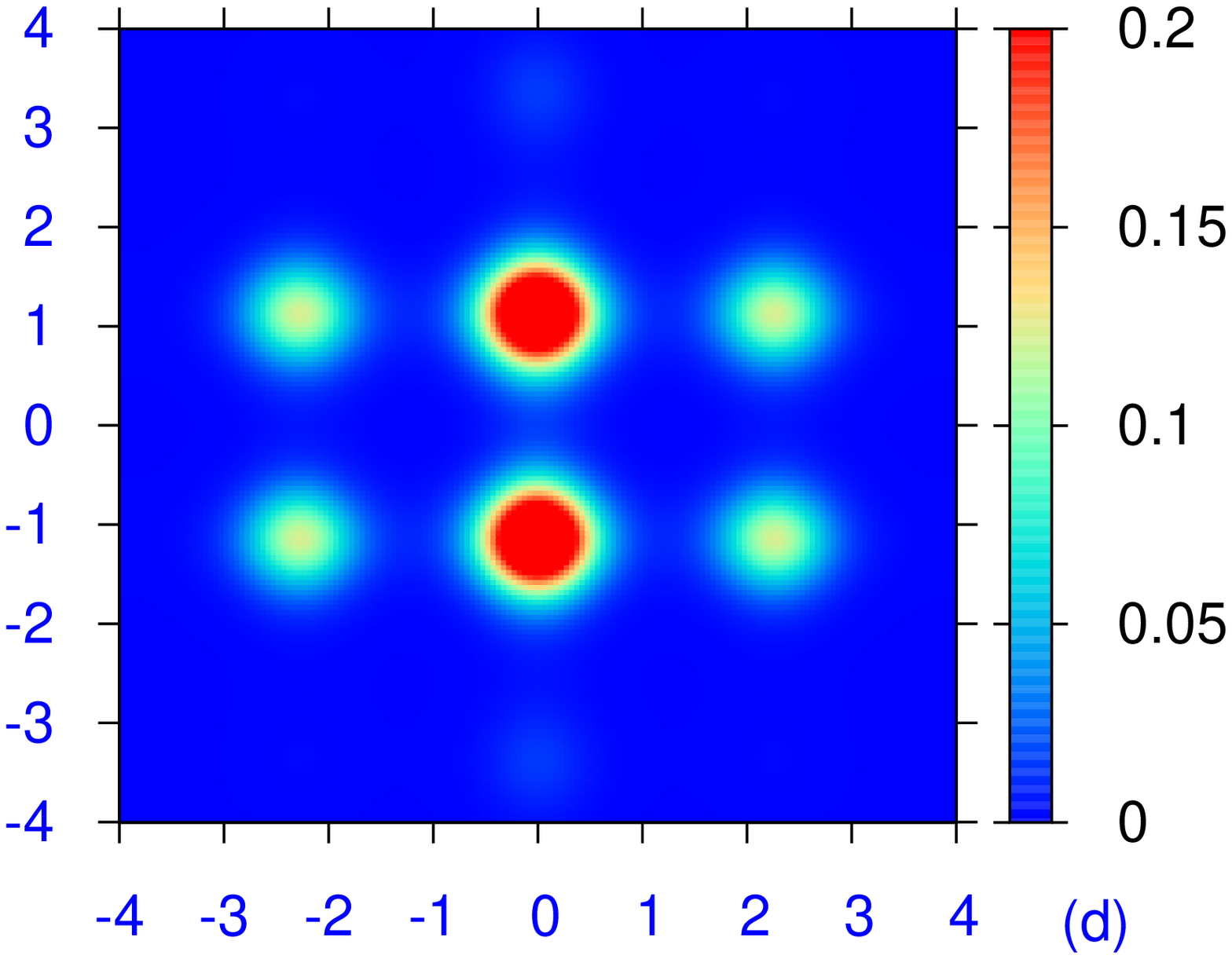}
\end{center}
\caption{(Color online)  
(a) Density  $\phi^2(x,y)$  
of a disk-shaped  BEC 
 from a numerical solution of Eq. 
(\ref{gp2d})  with (b) its contour plot 
for  potential (\ref{p2}) along $x$ direction and 
potential (\ref{p1}) along $y$ direction 
for  $g=0$.  
(c) Same as (a) but with $g=2$. (d) Same as (b) but with $g=2.$
All quantities are in  oscillator 
units.}
\label{fig3}
\end{figure}

%The shape of the localized state then becomes a quasi-Gaussian for 
%potential (\ref{p2}) with $g= 0$ and we compare our numerical results 
%with the variational result in this case.

In Figs. \ref{fig2} (a) and (b) 
we plot 
the results for density 
$\phi^2(x,y)$ vs. $x$ and $y$ from numerical and 
variational solutions of Eqs. (\ref{gp2d}) and  (\ref{p2}) for   
  $g=0$, respectively. 
(The $g=0$ case is trivial as the 2D and 
3D Eqs. (\ref{gp2d}) and (\ref{gp3d}) decouple into two and 
three 1D equations, respectively. Nevertheless, this case allows us to test 
carefully the numerical programs by comparing the variational result with the
two numerical results obtained by real- and imaginary-time propagation.)
The 
variational width ($w= 0.4864$) obtained from a solution of Eq. 
(\ref{cf}) produced the density  in good agreement with the 
numerical density  in Fig. \ref{fig2} (a). In Fig. \ref{fig2} (c) we 
plot the results for density 
$\phi^2(x,y)$ vs. $x$ and $y$ from a numerical solution
of Eqs. (\ref{gp2d}) and
(\ref{p2}) for the same set of parameters as in Fig. \ref{fig2} (a) but 
with $g=2$. (As in 1D \cite{adhikari,modugno2}, repulsive nonlinearity $g$ 
has a 
strong effect on localization.   As $g$  is increased to a small 
positive  value,
the 
localized states occupy more and more lattice sites with larger spatial 
extension.) In  Fig. \ref{fig2} (d) we show a contour plot of the density 
shown in Fig. \ref{fig2} (c). 
Because of the nonlinear inter-atomic repulsion, the 
essentially single peak of Fig. \ref{fig2} (a) now transforms into a 
multi-peak structure, although the central peak is still the strongest. 
Finally, in  Figs. \ref{fig2} (e) and (f) we plot $\phi^2(x,y)$ 
vs. $x$ and $y$ and its contour plot 
from a numerical solution of Eqs. (\ref{gp2d}) and 
(\ref{p1}) for $g=2$. 
Due to a change in the nature of the potential from Eq. (\ref{p2}) (sine,
minimum of potential at origin)
 to Eq. (\ref{p1}) (cosine, maximum of potential at origin)  the 
central region has a minimum of density 
in this case and not a maximum as in Figs. 
\ref{fig2} (a) and (c). The difference in the structure of the  
BEC densities can clearly be seen from the respective contour plots.

\begin{figure}%[!ht]
\begin{center}
\includegraphics[width=\linewidth]{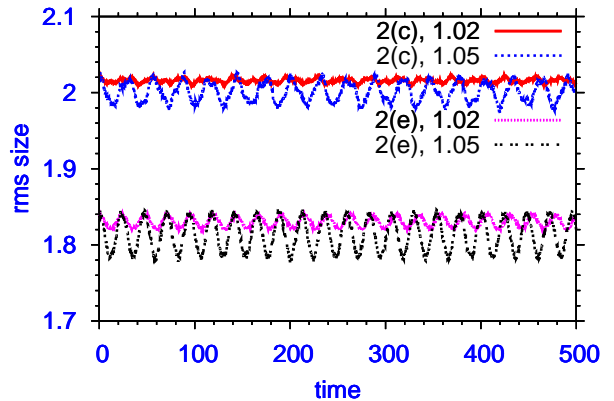}
\end{center}

\caption{(Color online) 
Time evolution of the rms sizes of the BECs  
depicted in Figs. 
\ref{fig2} (c)  and (e) 
as the potential is multiplied suddenly by a factor 
of 1.02 or 1.05. All quantities are in oscillator units.
}

\label{fig4}
\end{figure}

We also calculated the chemical potential of the states illustrated in 
Figs. \ref{fig2}. The numerical result for energy for potential 
(\ref{p2}) of Fig. \ref{fig2} (a) is 4.747, to be compared with the 
variational result of Eq. (\ref{energy}) 4.791, calculated with width 
$w= 0.4864$ obtained by solving Eq. (\ref{cf}).  This agreement between 
the numerical and variational results of the BEC density for potential 
(\ref{p2}) in Fig. \ref{fig2} (a), and of the respective energies, 
provides assurance about the accuracy of the numerical code used in 
simulation in our investigation. We also calculated the (numerical) 
chemical potential of the BEC density displayed in Figs. \ref{fig2} (c) 
($\mu =5.102$) and (e) ($\mu =5.103$). The two chemical potentials are 
practically the same. A similar finding was noted in the 1D case as the 
potential was changed from sine to cosine type \cite{adhikari}.

As we are considering two potentials (\ref{p1}) and (\ref{p2}) in 2D and 
3D it is possible to have a distinct  potential along each axis. We 
consider this possibility in the case of the localized state in a 
disk-shaped BEC where we take potential (\ref{p1}) along $x$ direction 
and potential (\ref{p2}) along $y$ direction with  $g=0$. The 
density  and the  contour plot of the BEC  
in this case are shown in Figs. \ref{fig3} 
(a) and (b). Just to illustrate the effect of a small 
nonlinearity on the localized 
state we repeated the numerical simulation in this case with a 
nonlinearity $g=2$. The density  and the corresponding contour plot 
of the resulting localized state  
are shown in Figs. \ref{fig3} (c) and (d). The density  in 
Fig. \ref{fig3} (c) is quite similar to that in Fig. \ref{fig3} (a), with the 
exception that the density  in Fig. \ref{fig3} (c)
extends over more sites due to 
the repulsive nonlinearity.

Next we explicitly demonstrate that the densities  shown in 
Figs. \ref{fig2}  corresponds to 
 stable states. For this we consider the 
time evolution of the BECs illustrated in Figs. \ref{fig2} (c) and
 (e) 
in a slightly altered potential, e.g., the potential obtained by 
multiplying the original potentials 
by a factor of 1.02 and 1.05.  The resultant  
rms sizes as 
calculated by the real-time propagation routine are plotted in Fig. 
\ref{fig4}. The curves in Fig. \ref{fig4} are labeled by Fig. 
\ref{fig2} (c) or (e) 
and the factor 1.02 or 1.05 which multiplies the 
potential.  In all cases the rms sizes execute breathing oscillation 
over a long period of time as shown in Fig. \ref{fig4} and this demonstrates
the stability of the localized  states.

\begin{figure}%[!ht]
\begin{center}
\includegraphics[width=.46\linewidth]{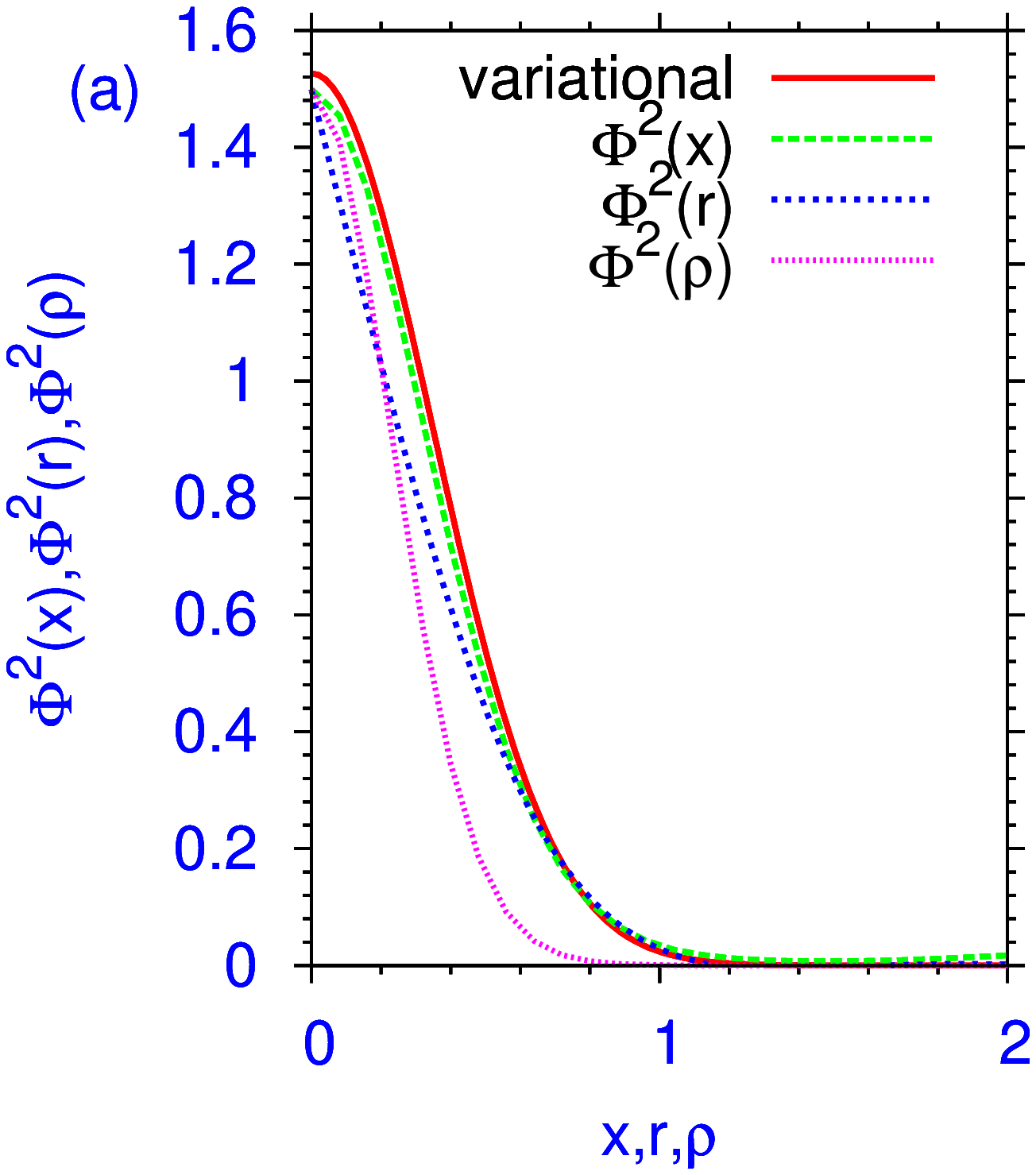}
\includegraphics[width=.46\linewidth]{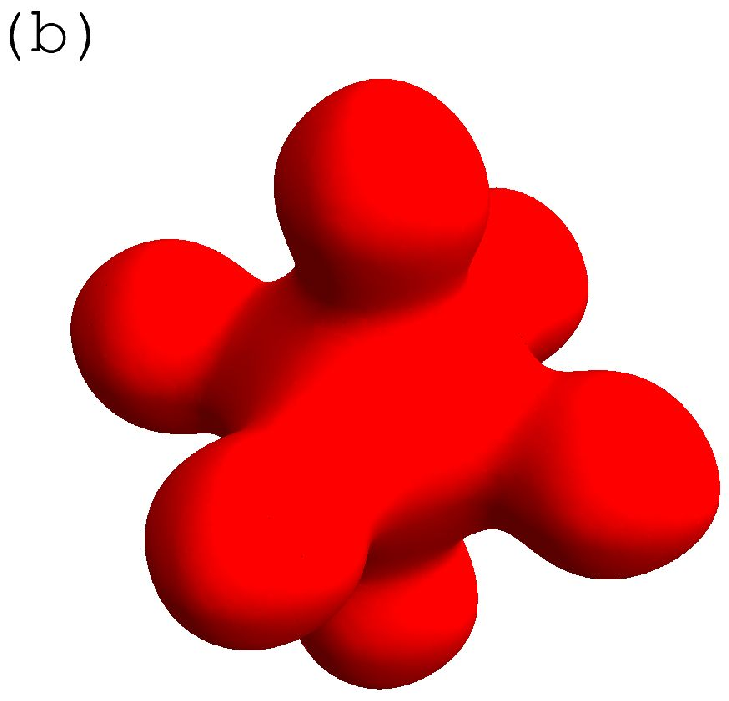}
\includegraphics[width=.46\linewidth]{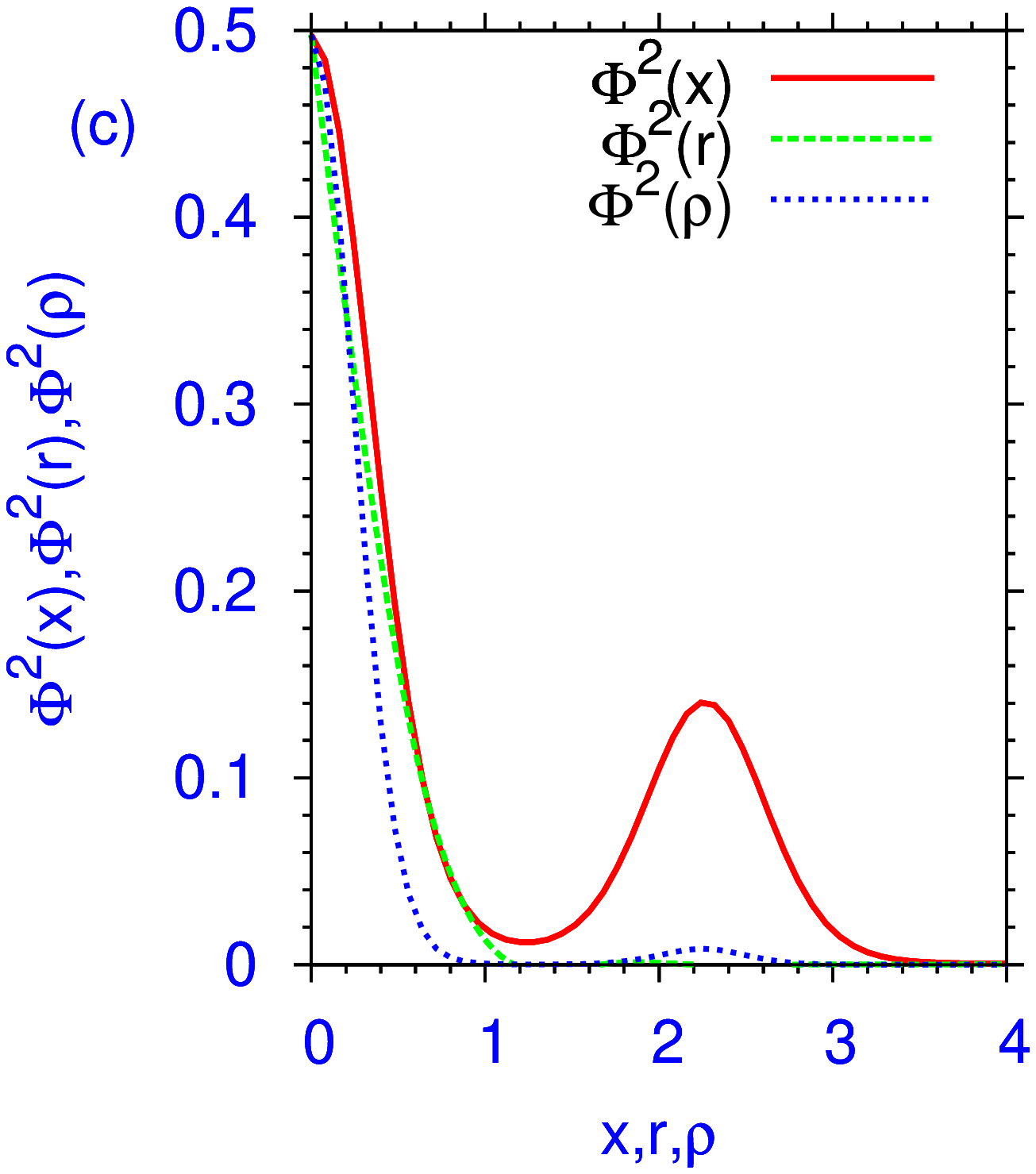}
\includegraphics[width=.46\linewidth]{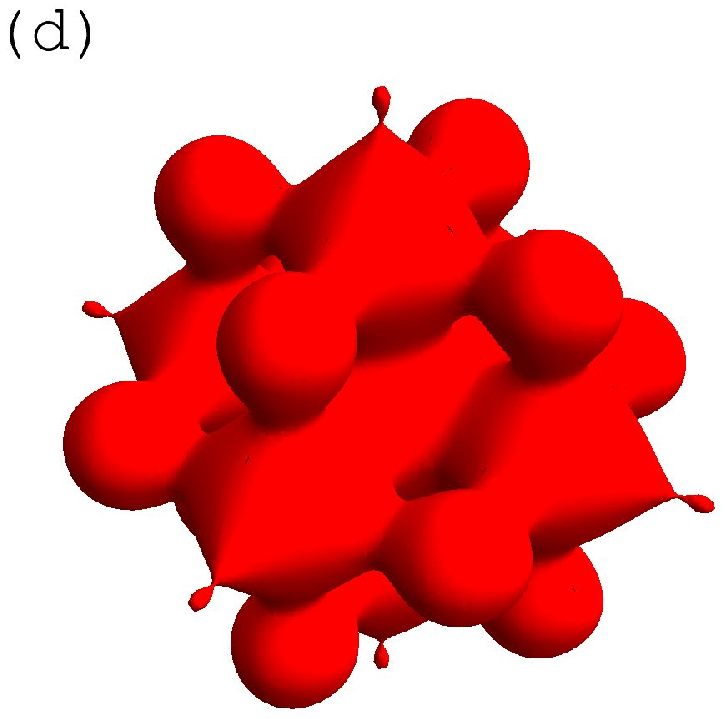}
\includegraphics[width=.46\linewidth]{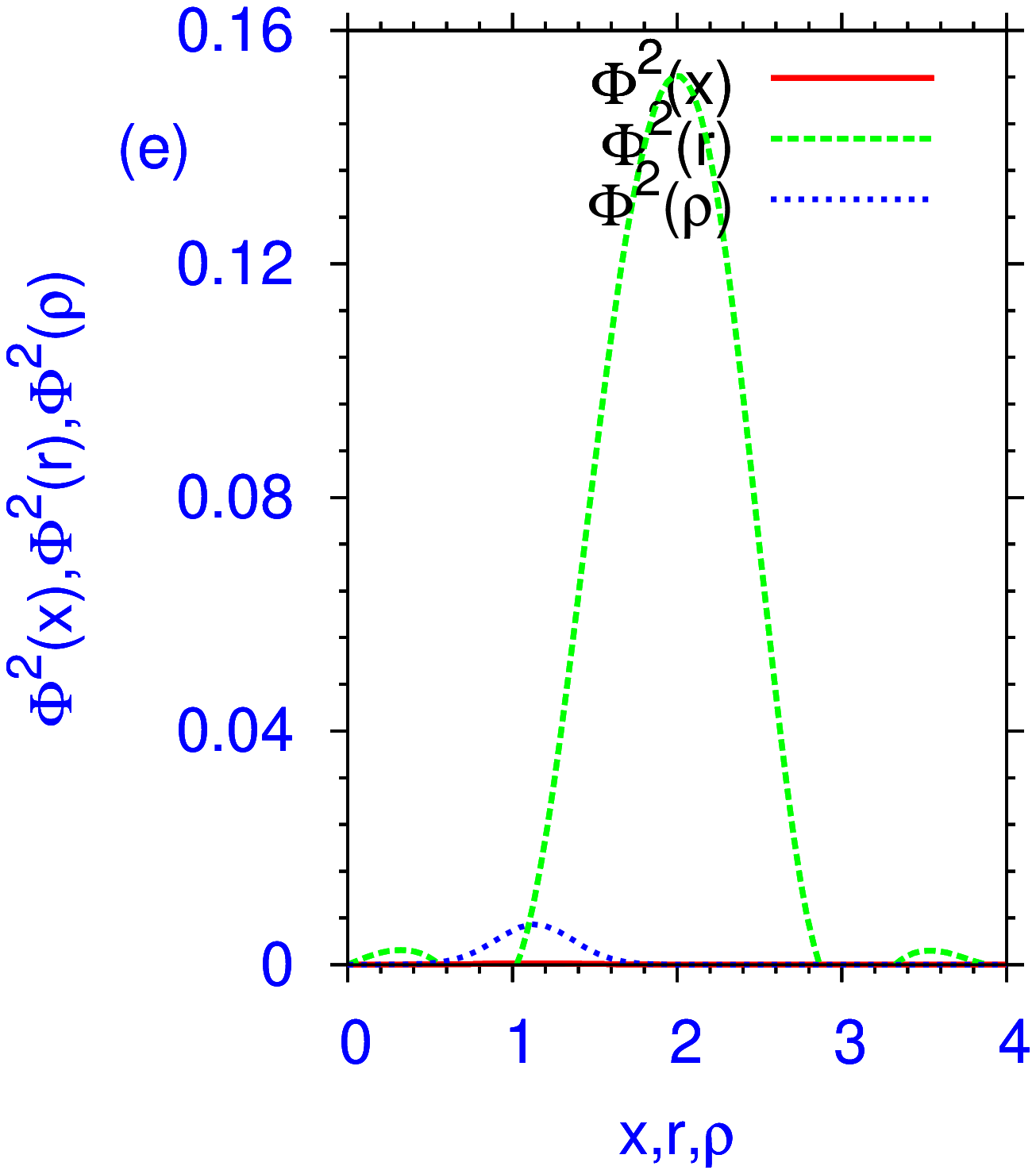}
\includegraphics[width=.46\linewidth]{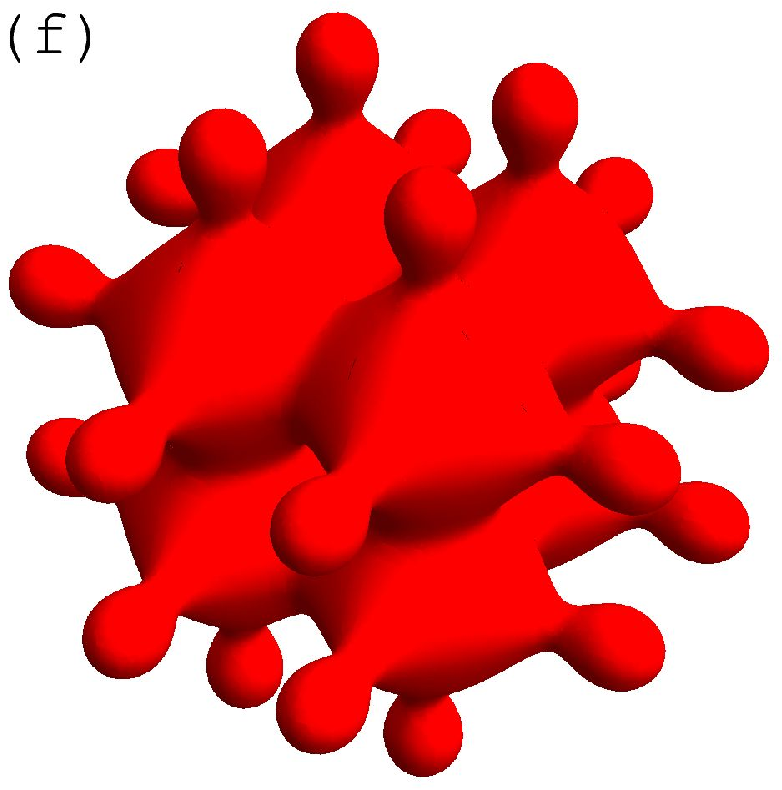}
\end{center}

\caption{(Color online) 
(a) Sections of  density  $\phi^2(x,y,z)$ for a 
3D  BEC  from a solution of Eq.
(\ref{gp3d})  with the (b)
3D
contour plot  of  density  $\phi^2(x,y,z)$
for potential (\ref{p2}) 
with  $ g=0$; 
(c) Same as (a) with $g=2$;
(d) Same as (b) with $g=2$;
(e) Same as (a) with potential (\ref{p1})
with the parameters  $\lambda_1=5, \lambda_2/\lambda_1=0.86, s_1=s_2=4$
and           $g=2$;
(f) Contour plot for (e).
The sections plotted in the left panel are $\Phi^2(x)\equiv 
\phi^2(x,0,0)$ vs.
$x$, $\Phi^2(\rho)\equiv \phi^2(0,\rho/\sqrt 2 ,\rho/\sqrt 2)$ vs. $\rho$, and 
$\Phi^2(r)\equiv \phi^2(r/\sqrt 3,r/\sqrt 3,r/\sqrt 3)$ vs. $r$.   
For potential (\ref{p2})
the variational result for $\Phi^2(r)$ vs. $r$ is also plotted. 
%In the contour  plots exhibited in (b), (d), and (f)
%the value of the  density  $\phi^2(x,y,z)$  at the boundary is 0.001.
}

\label{fig5}
\end{figure}

\begin{figure}%[!ht]
\begin{center}
\includegraphics[width=.48\linewidth]{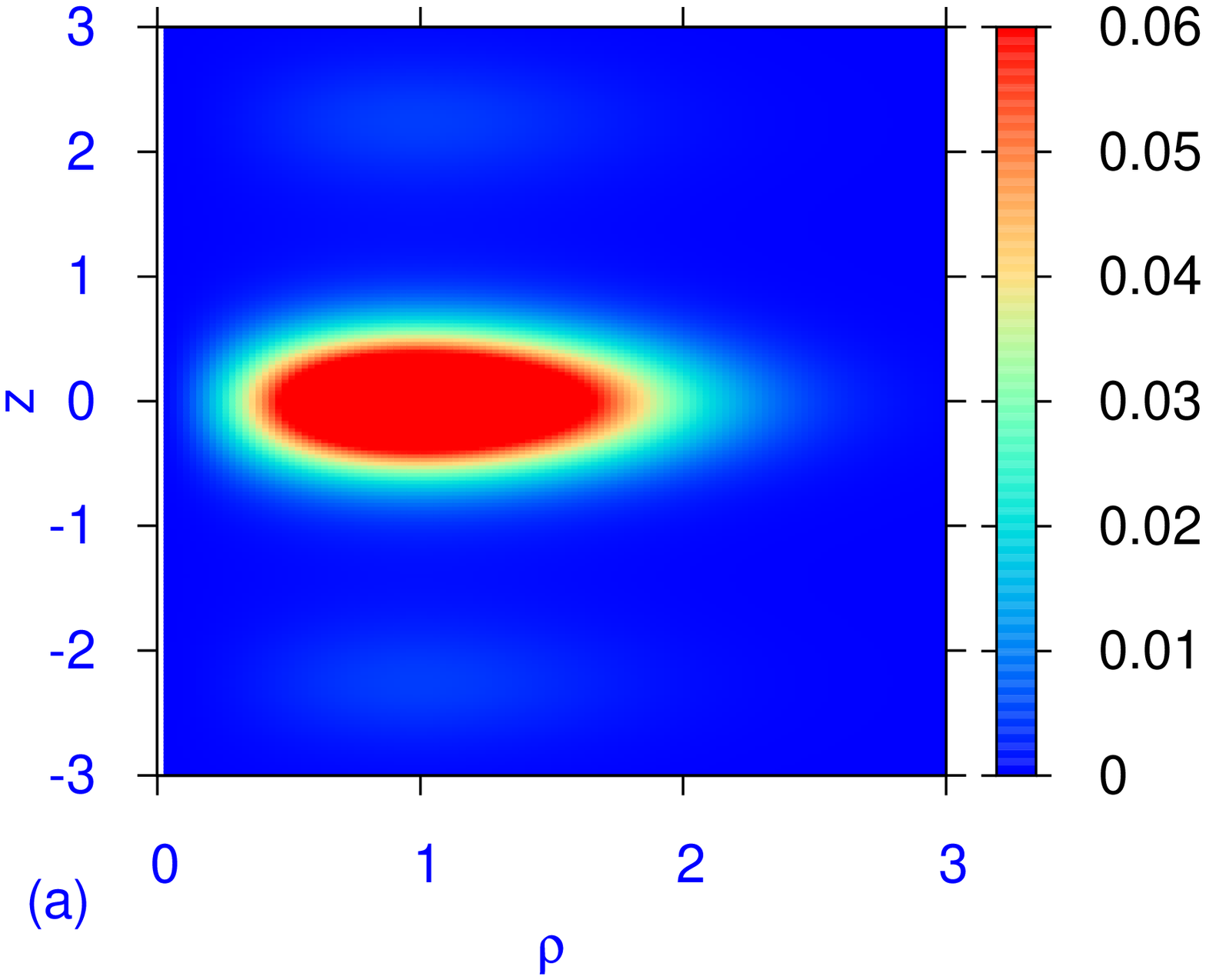}
\includegraphics[width=.48\linewidth]{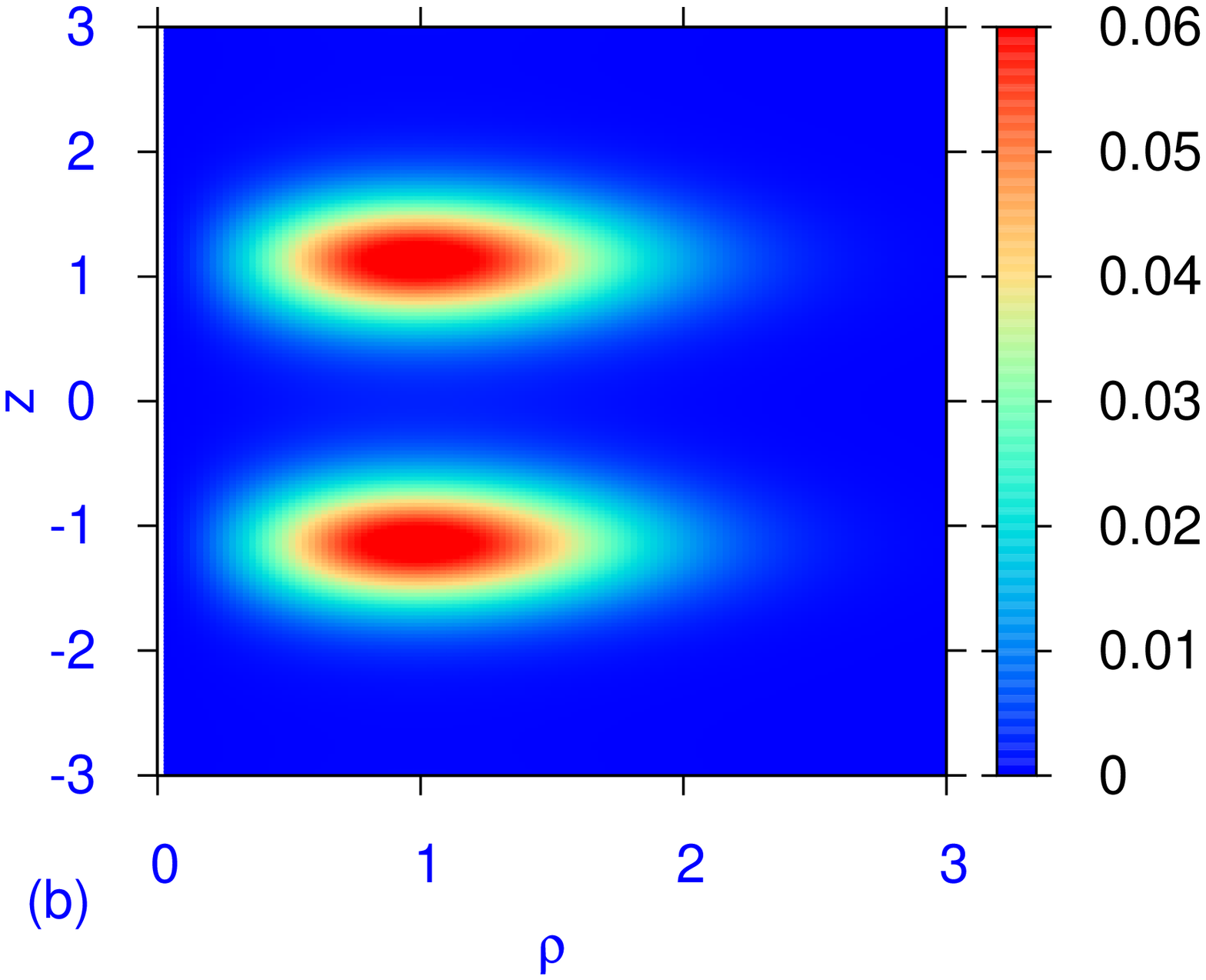}
\end{center}

\caption{(Color online) 
(a) 2D contour plot of the BEC  density  $\phi^2(\rho,z)$
for a localized 3D vortex in bichromatic OL potential
(\ref{p2}) 
with 
$g=5$,
obtained by solving Eq. (\ref{gp3dv}). The 
vortex core appears as the line $\rho =0$.
(b) Same as (a) in  bichromatic OL potential
(\ref{p1}). As potential (\ref{p1}) has a maximum at $z=0$ the matter 
density is zero there. 
}

\label{fig6}
\end{figure}

\subsection{3D Bichromatic Optical Lattice }

\label{IIIB}

Now we consider a few cases of the localized states in 3D 
as obtained from  a solution of Eq. (\ref{gp3d}) together with potential 
(\ref{p1}) or (\ref{p2}) along three orthogonal directions. First, 
as in 2D,  we consider the solution of Eq. (\ref{gp3d}) 
with potential (\ref{p2}) for  $g=0$. 
The result is illustrated in Figs. \ref{fig5} 
(a) and (b). In Fig.   \ref{fig5} (a)
 we plot three sections of the  density  $-$   $\Phi^2(x)
\equiv\phi^2(x,0,0)$ 
vs. $x$, $\Phi^2(r)\equiv \phi^2(r/\sqrt 3, r/\sqrt 3,r/\sqrt 3)$ vs. 
$r$, 
and $\Phi^2(\rho)\equiv \phi^2(0,\rho/\sqrt 2,\rho/\sqrt 2)$ vs. $\rho$
$-$
together with the variational result.
 Here $\Phi^2(x)$ corresponds to the density in the axial $x$ direction with 
polar angle $\theta=0$, 
$\Phi^2(r)$ that in the diagonal direction with polar angle $\theta=\pi/4$ 
and azimuthal angle $\varphi=\pi/4$, and 
$\Phi^2(\rho)$ that in the transverse direction with polar angle  
$\theta=\pi/2$ and azimuthal angle $\varphi=\pi/4$. 
In Fig.  \ref{fig5} (b) we show the 3D contour plot 
(obtained 
using Mathematica)
of the 
BEC  density  
showing 
the actual shape of the localized state. (The value of the 
 density   
at the boundary of the plot is 0.001 in Figs. \ref{fig5} (b), (d) and (f).)
As $g=0$ in Fig. \ref{fig5} (a), the 3D 
wave function  is trivial and decouples in the three directions. 
The variational result in this case is in good agreement with the 
numerical result for the  density  in the diagonal direction $-$ 
$\Phi^2(r)$.
Next we 
consider the nontrivial 3D case with $g=2$ for potential (\ref{p2}). 
In this case the 3D 
solution has no 1D counterpart.
In Figs. \ref{fig5} (c) and (d) we plot the  sections of the densities
and the 3D contour plot, respectively, 
 for potential (\ref{p2}) with  $g=2$.
The shape of the BEC  density  is similar to the 
$g=0$ case with a maximum at the origin, the only difference being that 
in Fig.  \ref{fig5} (c) the  density  
extends over a larger distance in space due to 
the repulsion introduced by a positive $g$ value. The  density   has 
secondary maxima in adjacent OL sites in this case. Finally, in Fig.  
\ref{fig5} (e) and (f) 
we show the results for the density  and its 3D contour plot, respectively,
with   
potential (\ref{p1})  and  $g=2$. Now 
the  density   has a minimum at the origin in contrast to the maxima in 
Figs. \ref{fig5} (a) and (c). Also, in   Figs. \ref{fig5}  (e) and (f)
 the  density  is zero 
along the three axes  (explicitly shown in $x$ direction with $\Phi^2(x)=0$).

Next we consider the localization in a 3D bichromatic OL potential  of a 
noninteracting BEC vortex
 with $g=5$ rotating around the $z$ 
direction with unit angular momentum from a numerical solution of 
Eq. (\ref{gp3dv}). In this case we consider  
potentials (\ref{p2}) and (\ref{p1})
along the $z$ direction.   
For a localized 3D vortex state rotating around the $z$ direction and 
described by Eq. (\ref{gp3dv}) under the action of the bichromatic OL 
potentials  (\ref{p2}) and (\ref{p1}), 
a vortex core of zero density passing 
through the origin develops along the $z$ axis. To illustrate the vortex 
state we solve Eq. (\ref{gp3dv}) with potentials   (\ref{p2}) 
and (\ref{p1}) for  $ g=5.$ In 
Figs. \ref{fig6} (a) and (b) 
we show the contour plot of the density 
$\phi(\rho,z)$ for potentials  (\ref{p2}) and (\ref{p1}), respectively.
For potential (\ref{p1}), in addition to the vortex core along the $z$ 
axis, the density is also zero along the $z=0$ line.  

\begin{figure}%[!ht]
\begin{center}
\includegraphics[width=\linewidth]{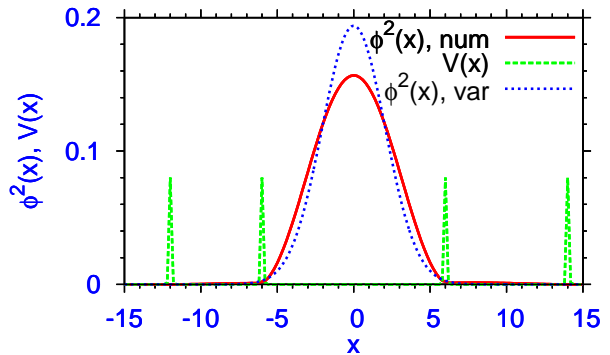}
\end{center}

\caption{(Color online)
The numerical (num) 
and
variational (var) 
BEC  density   $\phi^2(x)$ vs. $x$ for the random 
potential (\ref{randpot})
with $B_i=8, c_i=100, x_{10}=-12, x_{20}=-6, x_{30}=6, x_{40}=14$ and $g=0$. 
  The potential $V(x)$ is also plotted in arbitrary units. 
}

\label{fig7}
\end{figure}

\subsection{Random Spike Potential} 

\label{IIIC}

Now we consider the localization of a BEC in the 1D random potential 
(\ref{randpot}).  For   small values of the amplitude $B_i$, 
 a large number $S$ of spikes is needed for a good localization. However 
we consider the minimum number $S (=4)$ of spikes for localization. For 
a robust localization with $S=4$, we take $B_i=8$ and $c_i=100$ 
corresponding to a Gaussian spike of very small width. Next we had to 
choose the random positions $\beta_{i}$. Different sets of unevenly distributed 
$\beta_{i}$ produced localization and here we take 
$\beta_{i}=-12,-6,6,14$ for $i=1,2,3,4$.  This choice of the points sets 
the center of the localized BEC approximately at $x=0$. In Fig. 
\ref{fig7} we plot the localized BEC  density  $\phi^2(x)$ vs. $x$ for 
nonlinearity $g=0$. The variational and numerical chemical potentials are 
0.03739 and 0.03093, respectively. 
The plot of the potential (in arbitrary units) is 
also shown in Fig. \ref{fig7}.  
With this potential the localization is destroyed for a 
very small nonlinearity $g (\sim 1)$, unless the number of spikes is 
increased and we do not consider the localization of an interacting BEC 
here.

\section{SUMMARY}

In this paper, using the numerical solution \cite{CPC} of the GP 
equation \cite{GP} in 2D and 3D, we studied the localization 
\cite{roati,billy} of a noninteracting and weakly interacting BEC in a 
quasi-periodic bichromatic OL potential along different axes. We 
considered two analytical forms (sine and cosine) of the OL potential, 
e.g., (\ref{p1}) and (\ref{p2}) and  considered  the same or different 
potentials along different axes. First we consider the localization of a 
disk-shaped BEC  with a small repulsive nonlinearity under 
the action of the bichromatic OL potential (\ref{p1}) or 
(\ref{p2}). 
We considered the localization 
of a full 3D BEC under the action of bichromatic OL potentials
(\ref{p1}) or 
(\ref{p2}) along the three axes  
with or without a small nonlinear atomic interaction. The increase of 
nonlinearity destroys localization in all cases considered 
\cite{adhikari,dnlse,modugno2}. We also considered the localization 
of a noninteracting BEC vortex of unit angular momentum with 
a 3D 
bichromatic OL potential (\ref{p2}) or (\ref{p1}) along the axial 
direction and harmonic traps along the transverse 
radial directions. The clear stable vortex cores 
in 3D as 
shown in Figs. \ref{fig6} (a) and (b) are the most important findings of 
this paper. 
{{Finally, we showed that one can have localized noninteracting 
BEC states 
under the action of random potentials taken in the form of repulsive spikes 
randomly distributed in space. All these localized states are found to 
be dynamically stable.
}}

We hope that the present work will motivate new studies, specially 
experimental ones, on localization in a more realistic 2D and 3D BEC 
under the action of bichromatic OL potentials along different axes. 
Specially challenging is the localization of a singly-quantized vortex 
BEC in 3D
 as predicted in this work. 
The consideration of a vortex lattice of BEC 
\cite{ketterle} in bichromatic OL potentials is of great interest also. 
It remains to be seen if one has vortex pinning in a rotating BEC as 
observed in a monochromatic OL potential superposed on a 
harmonic trap \cite{cornell}. The effect of a bichromatic OL potential
on a two-species mixture of BECs \cite{two} is also of interest.
In the bichromatic OL potential considered in this paper, the disorder 
term is taken in the form of a periodic potential with an incommensurate  wave 
length. 
We also considered a disordered potential in the form of narrow 
identical repulsive spikes distributed randomly in space. 
 Further work 
need be done to study the effect of different types 
disorder terms. The 2D and 
3D localization of BEC under the action of bichromatic OL potentials 
and of a random potential 
$-$ as 
predicted in this paper $-$ may find application in other contexts in 2D 
and 3D, for example, in the localization of electron waves in 2D and 3D 
lattice as well as in the localization of 
electromagnetic waves, or sound waves in the presence 
of disorder.

\label{V}

\acknowledgments

FAPESP and  CNPq (Brazil)
provided partial support.

\end{document}